\documentclass[a4paper,12pt]{article}
\pdfoutput=1

\usepackage[usenames]{color}
\usepackage[colorlinks=true
,urlcolor=blue
,anchorcolor=blue
,citecolor=green
,filecolor=blue
,linkcolor=blue
,menucolor=blue
,pagecolor=blue
,linktocpage=true
]{hyperref}
\usepackage[centertags]{amsmath}
\usepackage{amsfonts}
\usepackage{amssymb}
\usepackage{bbm}
\usepackage{amsthm}
\usepackage{newlfont}
\usepackage[a4paper]{geometry}
\usepackage{graphicx}
\usepackage{dsfont}
\usepackage{feynmf}
\usepackage{cleveref}
\usepackage{array}

\newcommand{\tlambda}{\tilde{\lambda}}
\newcommand{\hlambda}{\hat{\lambda}}

\newcommand{\YM}{\mathrm{YM}}

\newcommand{\qb}{\mathrm{q.b.}}

\newcommand{\dho}{\partial}

\newcommand{\ed}{\,.}
\newcommand{\ec}{\,,}
\newcommand{\ecq}{\ec\quad}

\newcommand{\cA}{\ensuremath{\mathcal{A}}}

\newcommand{\cD}{\ensuremath{\mathcal{D}}}

\newcommand{\cN}{\ensuremath{\mathcal{N}}}
\newcommand{\cO}{\ensuremath{\mathcal{O}}}
\newcommand{\cP}{\ensuremath{\mathcal{P}}}

\begin{document}

\title{
\vspace{-20mm}
\hfill{\small \tt WIS/16/12-NOV-DPPA}\vskip 5pt
Correlators of Large $N$ Fermionic Chern-Simons Vector Models}
\author{Guy Gur-Ari and Ran Yacoby\\\\
{\it Department of Particle Physics and Astrophysics}\\
{\it Weizmann Institute of Science, Rehovot 76100, Israel}\\
{\small{\tt e-mails~:~Guy.GurAri,~Ran.Yacoby@weizmann.ac.il}}}

\maketitle
\vspace{-7mm}
\begin{abstract}
We consider the large $N$ limit of three-dimensional $U(N)_k$ Chern-Simons
theory coupled to a Dirac fermion in the fundamental representation. In this
limit, we compute several correlators to all orders in the `t~Hooft coupling
$\lambda \equiv N/k$.  It was suggested recently that this theory is dual to
the Legendre-transformed theory of scalar fields coupled to Chern-Simons gauge
interactions.  Our results show that this duality holds for any value of the
`t~Hooft coupling, at least at the level of the planar 3-point functions. In
addition, we determine the sign in the duality transformation of the
Chern-Simons level, as well as the relation between the ``triple-trace''
deformation which exists in the bosonic Chern-Simons theory and in the
Legendre-transformed fermionic theory.
\end{abstract}

\newpage

\tableofcontents

\setlength{\unitlength}{1mm}

\section{Introduction}

Let us consider three-dimensional free field theories with $N$ massless complex bosons $\phi^i$ or Dirac fermions $\psi^i$. These conformal field theories are known as the ``free $U(N)$ vector models'', and we will restrict ourselves to their $U(N)$ singlet sector. The closely-related ``critical $U(N)$ vector models'' are interacting conformal field theories that can be defined (in the large $N$ limit) by performing a Legendre transform of the free theories with respect to the scalar primary operators $\bar{\phi}^i\phi_i$ or $\bar{\psi}^i\psi_i$. Because we gauged the $U(N)$ symmetry, these four theories can be naturally coupled to a gauge field with Chern-Simons interactions at level $k$. The singlet sector of the vector models is recovered by taking $k$ to infinity.

In the large $N$ limit with fixed `t~Hooft coupling $\lambda=N/k$, these theories have remarkably simple properties.
First, both the bosonic and fermionic theories are conformal\footnote{At finite $N$ the fermionic theory is still conformal. The bosonic theory has a possible deformation $(\bar{\phi}^i\phi_i)^3$ that is marginal in the planar limit, but not at finite $N$. At weak coupling one may nevertheless reach a fixed point by tuning the coupling of this deformation \cite{Aharony:2011jz}.}. Second, the full spectrum of gauge-invariant local operators in these theories is independent of $\lambda$ at large $N$ \cite{Aharony:2011jz,Giombi:2011kc}. The structure of the planar $3$-point functions of these operators is also highly constrained \cite{Maldacena:2012sf}. In particular, \cite{Maldacena:2012sf} determined all the planar $3$-point functions of primary operators in terms of two parameters\footnote{In the bosonic and critical-fermionic Chern-Simons theories there is a third parameter, corresponding to the deformation mentioned above, that only affects the $3$-point function of the primary scalar operator.} $\tilde{N}$ and $\tilde{\lambda}$, which are some functions of $N$ and $\lambda$.

The results of \cite{Maldacena:2012sf} also imply that the planar 3-point functions of the free fermionic vector model turn into those of the critical bosonic model, when one changes the coupling $\tilde{\lambda}$ from zero to infinity. A similar relation holds between the free bosonic and critical fermionic vector models. This three dimensional ``bosonization'' was first conjectured in \cite{Giombi:2011kc} by considering the gravity duals of these theories. In particular, it implies that (at least at the level of the planar $3$-point functions) the $U(N_f)_{k_f}$ fermionic Chern-Simons vector model is equivalent to $U(N_b)_{k_b}$ critical-bosonic Chern-Simons vector model, for some integers $N_f,k_f$ and $N_b,k_b$.

In \cite{Aharony:2012nh}, the parameters $\tilde{N}$ and $\tilde{\lambda}$ of the bosonic Chern-Simons theories were determined exactly as functions of $N$ and $\lambda$. Using this result it was found that, in the large $N$ limit, the $U(N)_{k}$ critical-bosonic Chern-Simons theory is equivalent to the $U(|k|-N)_{\pm k}$ fermionic theory.  Note that the pure Chern-Simons theories on both sides of this relation are equivalent by level-rank duality
\cite{Naculich:1990pa,Camperi:1990dk,Mlawer:1990uv}, which is valid for any $N$ and $k$. The duality takes a more familiar form when written in terms of a shifted level $k_{\YM}$, defined by $k=k_{\YM}+\mathrm{sign}(k_{\YM})N$; it is given by $U(N)_{k_{\YM}}\leftrightarrow U(|k_{\YM}|)_{N}$ (up to the level's sign), where the subscript denotes the value of the shifted level. This is an important consistency check on the proposed duality in the large $N$ limit; in some computations, such as correlation functions of Wilson lines, the leading contributions are of order $N^2$ and are completely determined by the pure Chern-Simons contribution.

The duality relation above was found by taking the strong coupling limit of the exact expressions for $\tilde{N}$ and $\tilde{\lambda}$ in the critical bosonic theory, and comparing the result to the same parameters in the weakly coupled fermionic theory (which were determined perturbatively in \cite{Giombi:2011kc}). The relative sign between the Chern-Simons levels in this relation was not determined. In this work we repeat the computations of \cite{Aharony:2012nh} for the case of the fermionic Chern-Simons vector model. We find that the above duality holds for any value of the coupling (and not just in the weak coupling limit of the fermionic theory), and we determine the relative signs between the levels in the duality relation. Moreover, the bosonic (and critical-fermionic) theories contain a ``triple-trace'' deformation, which in the bosonic theory can be written as $\lambda_6^b (\bar{\phi}^i\phi_i)^3$. This deformation is exactly marginal at large $N$ \cite{Aharony:2011jz}, and we determine how it maps between the two theories.

Let us summarize our results. We find that the planar 3-point functions of the $U(N)_k$ Chern-Simons theory coupled to a fundamental scalar field in the critical fixed point, are equal to the same correlators in the $U(|k|-N)_{-k}$ Chern-Simons theory coupled to a fundamental fermion. This provides further evidence that these two theories are equivalent\footnote{At finite $N$, the conjectured duality must involve a half-integer shift of the Chern-Simons level of the fermionic theory due to the parity anomaly \cite{Aharony:2012nh}.}. The same duality map exists between the 3-point functions of the bosonic and critical fermionic Chern-Simons theories. In addition, the ``triple-trace'' deformations of the latter two theories are related by
\begin{align}
\lambda_6^b = 8\pi^2(1-|\lambda_f|)^2\left( 3 - 8\pi \lambda_f \lambda_6^f \right) \ed
\end{align}

The paper is organized as follows. In section \ref{conv} we introduce our conventions and methods. In sections \ref{vertices} and \ref{corr} we present the detailed computations of correlation functions in the fermionic theory. The reader interested in the final results can skip directly to section \ref{analysis} in which we use our results to determine the duality map between the fermion and boson theories. In section \ref{crit} we present the results for the critical fermionic theory. We conclude with a discussion in section \ref{conc}. Appendix \ref{positivity} includes a discussion of reflection positivity in theories with a non-real action.

\section{Fermionic Chern-Simons Vector Model}
\label{conv}

Consider the theory of a Dirac fermion $\psi$ in the fundamental representation of $U(N)$, coupled to gauge bosons $A_\mu$ with Chern-Simons interactions at level $k$ in three Euclidean dimensions. The action is\footnote{The covariant derivative is $\cD_{\mu} \equiv \dho_{\mu} + A_{\mu}^a T^a$. The generators $T^a$ are anti-hermitian, in the fundamental representation of $U(N)$, and normalized such that $\text{tr}(T^aT^b) = -\frac{1}{2}\delta^{ab}$. The Dirac matrices are the ordinary Pauli matrices $\gamma^{\mu}=\sigma^{\mu}$, $\mu=1,2,3$, and $\bar{\psi} = \psi^\dagger$.}
\begin{align}
  S &= - \frac{i k}{8\pi} \int \! d^3x \, \epsilon^{\mu\nu\rho}
  \left(A^a_{\mu}\dho_{\nu}A^a_{\rho}
  + \frac{1}{3}f^{abc}A^a_{\mu}A^b_{\nu}A^c_{\rho}\right) +  \int \! d^3x \,\bar{\psi}\gamma^{\mu}\cD_{\mu}\psi \ed
\label{action}
\end{align}

We work in the `t~Hooft large $N$ limit, keeping $\lambda=\frac{N}{k}$ fixed. The normalization of the Chern-Simons action \eqref{action} is such that the theory is gauge invariant if $k \in \mathds{Z}$, up to a $\pm \frac{1}{2}$ shift due to the parity anomaly \cite{Niemi:1983rq,Redlich:1983kn,Redlich:1983dv}; this shift will not matter to us since we work in the large $k$ limit. We use light-cone gauge, $A_-=0$, where the light-cone coordinates are defined by $x^{\pm}=x_{\mp}=(x^1\pm ix^2)/\sqrt{2}$. As was first noticed in \cite{Giombi:2011kc}, this gauge choice leads to great simplifications in perturbative calculations, since the self-interaction of the gauge bosons vanishes.

In this gauge the gluon propagator, which is exact in the planar limit, evaluates to
\begin{align}
  \langle  A_\mu^a(-p) A_\nu^b(q) \rangle  &= G_{\nu\mu}(p) \delta^{ab}
  \cdot (2\pi)^3 \delta^3(q-p) \ec \notag \\
  G_{+3}(p) &= - G_{3+}(p)
  = \frac{4\pi i}{k} \frac{1}{p^+} \ec
  \label{gluon-prop}
\end{align}
and the other components of $G$ vanish. The exact planar fermion propagator in light-cone gauge was computed in \cite{Giombi:2011kc} by summing over rainbow diagrams, and was found to be
\begin{gather}
\langle \psi_i(p) \bar{\psi}^j(-q) \rangle = \delta_i^j S(p) \cdot (2\pi)^3 \delta^3(p-q) \ec \\
S(p) \equiv \frac{ -i\gamma^{\mu}p_{\mu} + i\lambda^2\gamma^+ p^- + \lambda p_s}{p^2} \ec
\label{fermion-prop}
\end{gather}
where $p_s^2 = 2p^+p^- = p_1^2 + p_2^2$.

As in \cite{Aharony:2012nh} we use dimensional regularization in the $x^3$ direction and a cutoff $\Lambda$ on the momenta in the radial direction of the $x^1-x^2$ plane. This regularization, as well as the light-cone gauge-fixing condition, preserve an $SO(2)$ rotation symmetry in the $x^1-x^2$ plane. While this cutoff prescription breaks Lorentz invariance (as does our choice of gauge), conformal invariance and gauge invariance, these symmetries are restored by demanding that the theory is conformally-invariant in the continuum limit.

The spectrum of primary operators of this theory contains a single primary operator $J^{(s)}$ for each integer spin $s\ge 1$, with dimension $\Delta=s+1+O(1/N)$, and a scalar $J^{(0)}$ of dimension $\Delta=2+O(1/N)$ \cite{Giombi:2011kc}. All other primaries are obtained by products of these ``single-trace'' operators, and at large $N$ their dimension is just the sum of the dimensions of the individual single-trace operators in the product. In this paper we will compute correlation functions of the spin 0 and spin 1 operators
\begin{align}
J^{(0)} &= \bar{\psi}^i\psi_i \ec \\
J^{(1)}_{\mu} &= i\bar{\psi}^i \gamma_{\mu} \psi_i \ed
\end{align}
As we explain in appendix \ref{positivity}, the factor of $i$ is included in $J^{(1)}$ to make its 2-point function positive-definite (as it is in the critical bosonic theory defined in \cite{Aharony:2012nh}).

\section{Exact Vertices}
\label{vertices}

In this section we compute the exact vertices of $J^{(0)}$, $J^{(1)}_+$ and $J^{(1)}_-$, which are defined by the correlators
\begin{align}
  \langle J^{(0)}(-q) \psi_i(k) \bar{\psi}^j(-p) \rangle   &=
  V(q,p) \delta_i^j \cdot (2\pi)^3 \delta^{(3)}(q+p-k)
  \ec \label{J0vertex} \\
  \langle J^{(1)}_\pm(-q) \psi_i(k) \bar{\psi}^j(-p) \rangle &=
  V_\pm(q,p) \delta_i^j \cdot (2\pi)^3 \delta^{(3)}(q+p-k)
  \label{Jpmvertex} \ed
\end{align}
Here we suppress the spinor indices on fermion fields and on $V$ and $V_{\pm}$, which are matrices in spinor space. Moreover, in the above correlators we amputate the external fermion propagators.

\begin{figure}[!ht]
\centering
  \includegraphics[width=0.9\textwidth]{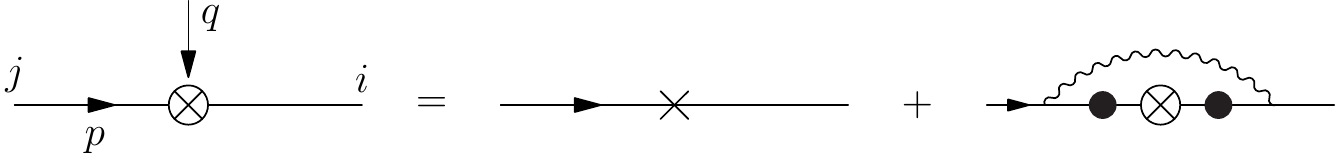}
  \caption{Schematic form of the Schwinger-Dyson equations for the vertices. The circled cross denotes an insertion of an exact $J^{(0)}$ or $J^{(1)}$ vertex. The un-circled cross denotes an insertion of the free vertex. A filled circle denotes the exact fermion propagator.}
  \label{fig:Fbs}
\end{figure}

The computation is done by solving the large $N$ Schwinger-Dyson equations for $V$ and $V_{\pm}$ in light-cone gauge. The equations are shown in Figure \ref{fig:Fbs}. We set $q^{\pm}=0$, which leads to simplifications in the calculations. These exact vertices will be used in section \ref{corr} as building blocks for the gauge invariant correlation functions of $J^{(0)}$ and $J^{(1)}$.

\subsection{$J^{(0)}$ Vertex}

Let us first consider the scalar operator vertex \eqref{J0vertex}. The bootstrap equation in Figure \ref{fig:Fbs}, with the circled cross denoting $V(q,p)$, is given by
\begin{align}
V(q,p) = 1 - 2\pi i \lambda \int \frac{d^3k}{(2\pi)^3} \gamma^{[3|}S(k+q)V(q,k)S(k)\gamma^{|+]}\frac{1}{(k-p)^+} \ed
\label{eq:V0bs}
\end{align}
The vertex $V(q,p)$ is a 2-by-2 matrix in spinor space. It can be expanded in a complete set of matrices consisting of $\gamma^{\mu}$ and the $2\times 2$ identity matrix $I$ as
\begin{align}
  V = v_{\mu}\gamma^{\mu} + v_II \ed
  \label{V}
\end{align}
Note that a general 2-by-2 matrix $A = a_{\mu}\gamma^{\mu} + a_II$ satisfies $\gamma^{[3|}A\gamma^{|+]} = 2\left( a_I \gamma^+ - a_- I\right)$. It then follows from \eqref{eq:V0bs} that $v_-=v_3=0$. Plugging \eqref{V} and the fermion propagator \eqref{fermion-prop} into \eqref{eq:V0bs}, performing the Dirac algebra, and separating the $I$ and $\gamma^+$ components, we obtain
\begin{align}
  v_I(q,p) &= 1 - 4\pi i\lambda \int \! \frac{d^3k}{(2\pi)^3} \,
  \frac{v_I(q,k)\left( 2i\lambda k_s - q_3 \right)k^+
  + 2v_+(q,k)(k^+)^2 }{k^2(k+q)^2(k-p)^+}\ec \label{VIbs}
  \\
  v_+(q,p) &=  4\pi i\lambda \int \! \frac{d^3k}{(2\pi)^3} \,
  \frac{v_I(q,k)\left[ (1-2\lambda^2) k_s^2 +k_3(k_3+q_3) \right]
  + v_+(q,k)\left( q_3 + 2i\lambda k_s \right)k^+}{k^2(k+q)^2(k-p)^+} \label{Vpbs}\ed
\end{align}
From these equations and our choice of external momenta we can see that $V(q,p)$ can depend only on $p^{\pm}$ and $q_3$. Furthermore, from dimensional analysis and rotational invariance in the $x^1-x^2$ plane, $V(q,p)$ can be written as
\begin{gather}
V(q,p) = v_+(q,p)\gamma^+ + v_I(q,p) I = \frac{2 \lambda p_+}{p_s}g\left(\frac{2p_s}{|q_3|}\right) \gamma^+ + f\left(\frac{2p_s}{|q_3|}\right) I \ec
\label{V0fg}
\end{gather}
where the factors are for later convenience. Our remaining task is to compute $f$ and $g$.

Let us plug the ansatz \eqref{V0fg} in \eqref{VIbs},\eqref{Vpbs}. We perform the integral over $k_3$, and then over the polar angle in the $k^1-k^2$ plane\footnote{The polar coordinates $k_s,\theta$ are defined by $k^{\pm}=k_se^{i\theta}/\sqrt{2}$.}. We are left with the integral equations
\begin{align}
f\left(\frac{2p_s}{|q_3|}\right) &= 1 - 2i\lambda\int_{p_s}^{\Lambda}dk_s\frac{\left(2i\lambda k_s - q_3\right)f\left(\frac{2k_s}{|q_3|}\right) + 2 \lambda k_s g\left(\frac{2k_s}{|q_3|}\right)}{4k_s^2+q_3^2} \ec \\
\frac{2\lambda p_+}{p_s}g\left(\frac{2p_s}{|q_3|}\right) &= -\frac{2i\lambda}{p^+}\int_{0}^{p_s}dk_s\frac{2k_s^2\left(1-\lambda^2\right)f\left(\frac{2k_s}{|q_3|}\right) + \left(q_3+2i\lambda k_s\right) k_s \lambda g\left(\frac{2k_s}{|q_3|}\right)}{4k_s^2+q_3^2} \ed
\end{align}

It will be convenient to work with the dimensionless, parity-invariant\footnote{Our theory is invariant under a parity transformation combined with $\lambda \to -\lambda$.} variables
$y=\frac{2p_s}{|q_3|}$, $x=\frac{2k_s}{|q_3|}$, $\Lambda'=\frac{2\Lambda}{|q_3|}$, and $\hlambda=\lambda\cdot\mathrm{sign}(q_3)$. We obtain
\begin{align}
  f(y)   &= 1 - i\hlambda\int_y^{\Lambda'} \! dx \,
  \frac{ \hlambda x g(x) - ( 1 - i\hlambda x )f(x) }{1+x^2}
  \ec \label{eq:fInt}
  \\
  y g(y) &= -i\int_0^y \! dx \, \frac{ x^2(1-\hlambda^2)f(x)
  + \hlambda( 1 + i\hlambda x )x g(x)}{1+x^2}
  \ed \label{eq:gInt}
\end{align}
These equations can be solved, for instance by using Mathematica, and we find
\begin{align}
  f(y) &= \frac{1+e^{-2i\hlambda\arctan\left(y\right)}}
  {1+e^{-2i\hlambda\arctan\left(\Lambda'\right)}}
  \ec \label{V0f}
  \\
  g(y) &=
  \frac{1 - i\hlambda y - (1+i\hlambda y)
  e^{-2i\hlambda\arctan\left(y\right)}}
  {\hlambda y \left[ 1+e^{-2i\hlambda\arctan\left(\Lambda'\right)} \right]}
  \label{V0g} \ed
\end{align}

\subsection{$J^{(1)}$ Vertices}

The computation of the $J^{(1)}_\pm$ vertices \eqref{Jpmvertex} is similar to that of the $J^{(0)}$ vertex.
The Schwinger-Dyson equation for $J^\mu=i\bar{\psi}^i\gamma^{\mu}\psi_i$ is
\begin{align}
  V^{\mu}(q,p) = i\gamma^{\mu} - 2\pi i \lambda
  \int \! \frac{d^3k}{(2\pi)^3} \,
  \gamma^{[3|}S(k+q)V^{\mu}(q,k)S(k)\gamma^{|+]}\frac{1}{(k-p)^+} \ed
  \label{Jmubs}
\end{align}
As before, we choose $q^{\pm}=0$, and the vertices are independent of $p_3$. Based on similar considerations as in the $J^{(0)}$ case, we can write $V^+$ as
\begin{align}
  V^+(q,p) &= g^{(+)}(y) \gamma^+ + \frac{2p^+}{q} f^{(+)}(y) \ed
  \label{Vpfg}
\end{align}
Plugging this in the bootstrap integral \eqref{Jmubs}, and carrying out the $k_3$ and polar integrals as before, we find
\begin{align}
  f^{(+)}(y) &= i \hlambda \int_y^{\Lambda'} \! dx \,
  \frac{ (1-i\hlambda x)f^{(+)}(x)-g^{(+)}(x)}{1+x^2} \ec \\
  g^{(+)}(y) &= i + i \hlambda \int_y^{\Lambda'} \! dx \,
  \frac{x^2(1-\hlambda^2)f^{(+)}(x) + (1+i\hlambda x)g^{(+)}(x)}{1+x^2} \ed
\end{align}
The solution to these equations is
\begin{align}
  f^{(+)}(y) &= \frac{i}{2} \left[1 - e^{2 i \hlambda
  \left( \arctan(\Lambda')- \arctan(y)\right)}\right]
  \ec \label{Vpf}
  \\
  g^{(+)}(y) &= \frac{i}{2}
  \left[1 - i\hlambda y + (1 + i\hlambda y)
  e^{2 i \hlambda \left( \arctan(\Lambda') - \arctan(y)\right)} \right]
  \label{Vpg} \ed
\end{align}

Similarly, $V^-$ can be written as
\begin{align}
  V^-(q,p) &= i\gamma^- +  \frac{2(p^-)^2}{p_s^2} g^{(-)}(y) \gamma^+
  + \frac{2 p^-}{q_3}f^{(-)}(y) \ed
  \label{Vmfg}
\end{align}
The equations that result from \eqref{Jmubs} are
\begin{align}
  y^2 f^{(-)}(y) &= i\hlambda \int_0^y \! dx \, \frac{x^2 g^{(-)}(x) -
  (1-i\hlambda x)x^2 f^{(-)}(x) - i(1+\hlambda^2)x^2 + 2\hlambda x}{1+x^2}
  \ec \\
  y^2 g^{(-)}(y) &= -i\hlambda \int_0^y \! dx \, \frac{ (1-\hlambda^2)x^4
  f^{(-)}(x) + (1+i\hlambda x)x^2 g^{(-)}(x)
  - i(1-\hlambda^2)x^2(1-i\hlambda x)}{1+x^2} \ec
\end{align}
and their solution is
\begin{align}
  f^{(-)}(y) &= \frac{i}{2 y^2}
  \left[ 1 - 2i\hlambda y - e^{-2i\hlambda\arctan(y)} \right]
  \ec \label{Vmf}
  \\
  g^{(-)}(y) &= \frac{i}{2 y^2}
  \left[ (1+i\hlambda y) e^{-2i\hlambda\arctan(y)} - (1-i\hlambda y) \right]
  \ed \label{Vmg}
\end{align}

\section{Correlation Functions}
\label{corr}

\begin{figure}[!ht]
\centering
  \includegraphics[width=0.3\textwidth]{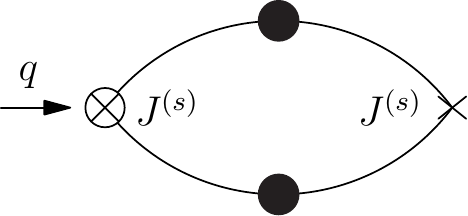}
  \caption{Diagrams contributing to $\langle J^{(s)} J^{(s)} \rangle$.}
  \label{fig:JsJs}
\end{figure}

\begin{figure}[!ht]
\centering
  \includegraphics[width=0.7\textwidth]{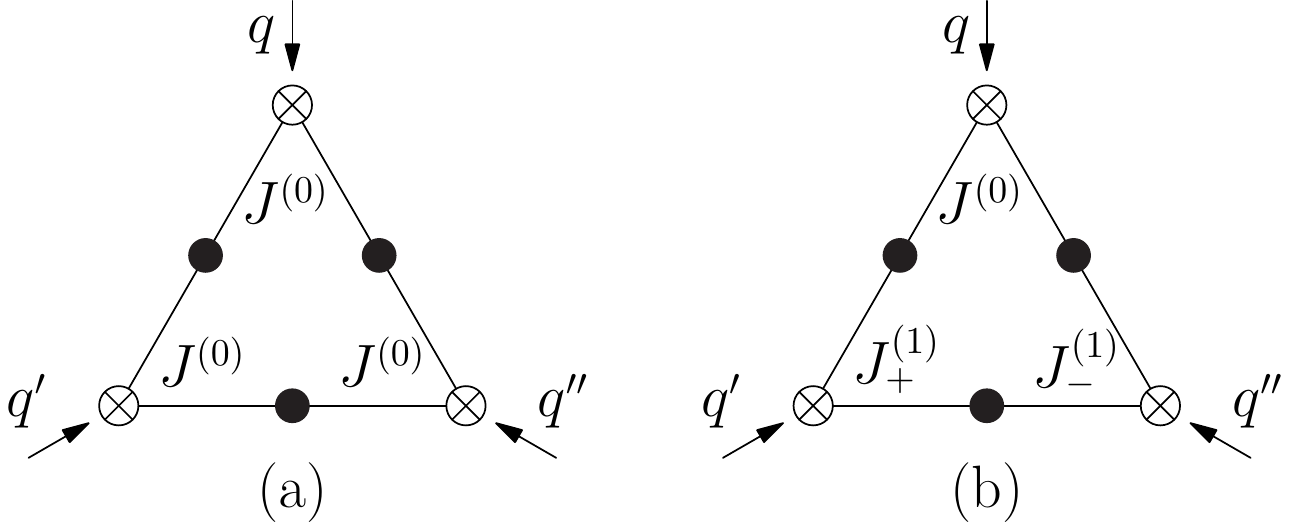}
  \caption{Diagrams contributing to $\langle J^{(0)} J^{(0)} J^{(0)} \rangle$
  and $\langle J^{(0)} J^{(1)}_+ J^{(1)}_- \rangle$.}
  \label{fig:JsJsJs}
\end{figure}

Using the results of the previous section, in this section we compute the exact planar 2-point and 3-point correlators of $J^{(0)}$ and $J^{(1)}$ with the external momenta all in the $x^3$ direction. The 2-point functions can be computed by the diagram in Figure \ref{fig:JsJs}. Note that only a single insertion of an exact vertex is required in the 2-point function to account for all the Feynman diagrams without any double-counting. The 3-point functions are obtained by computing the diagrams shown in Figure \ref{fig:JsJsJs}.

\subsection{2-point Functions}
\label{2pt}

\subsubsection{$\left<J^{(0)} J^{(0)}\right>$}
Using the exact $J^{(0)}$ vertex in \eqref{V0fg},\eqref{V0f},\eqref{V0g} and the fermion propagator \eqref{fermion-prop} the $2$-point function evaluates to
\begin{align}
  \langle J^{(0)}(-q) J^{(0)} \rangle &=
  -N \int \! \frac{d^3p}{(2\pi)^3} \, \text{Tr}\left[S(q+p)V(q,p)S(p)\right]
  \notag \\
  &\rightarrow
  -\frac{N\tan\!\left(\frac{\pi\lambda}{2}\right)}{4\pi\lambda}|q|
  + \frac{N}{2\pi}\Lambda \ec
  \label{J0J01}
\end{align}
where the overall minus is from the fermion loop, and in the last step we took $\Lambda \to \infty$, keeping divergent terms.

Let us denote the background source of $J^{(0)}$ as $\varphi$. The divergence in \eqref{J0J01} can be subtracted with a mass counterterm $\varphi^2$, whose coefficient is uniquely determined by demanding that the continuum theory be conformally-invariant. The renormalized $2$-point function is therefore
\begin{align}
  \langle J^{(0)}(-q) J^{(0)} \rangle =
  -\frac{N\tan\!\left(\frac{\pi\lambda}{2}\right)}{4\pi\lambda}|q| \ed
  \label{J0J0}
\end{align}
This result is also correct for general $q$. In appendix \ref{positivity} we verify that it is consistent with reflection positivity.

\subsubsection{$\left<J^{(1)} J^{(1)}\right>$}

Similarly, using the exact $J^{(1)}_+$ vertex \eqref{Vpfg},\eqref{Vpf},\eqref{Vpg} and the fermion propagator \eqref{fermion-prop}, we obtain\footnote{One could also use the exact $J^{(1)}_-$ vertex in this computation and get the same result.}
\begin{align}
  \langle J_-^{(1)}(-q) J_+^{(1)} \rangle &= -N i \int \! \frac{d^3p}{(2\pi)^3} \,
  \text{Tr}\left[S(p+q)V^+(q,p)S(p)\gamma^-\right]
  \notag \\
  &\rightarrow
  \frac{N i q (e^{i\pi\hlambda}-1)}{16\pi\lambda}
  + \frac{N}{4\pi}\Lambda \ed
  \label{JpJmraw}
\end{align}
The divergence can be removed by adding a mass term to the background gauge field $\cA_{\mu}$ that couples to $J_{\mu}^{(1)}$. The $\lambda$-odd part of \eqref{JpJmraw} is
\begin{align}
  - \frac{i N q}{8}
  \frac{\sin^2\!\left( \frac{\pi\lambda}{2} \right)}{\pi\lambda}
\ed \end{align}
This parity-odd\footnote{Here we are referring to pure parity transformation, without the additional $\lambda \to -\lambda$.} piece can come from a conformally-invariant contact term $\langle J_\mu(q) J_\nu \rangle \sim \epsilon_{\mu\nu\rho} q^\rho$, which corresponds to the appearance of a Chern-Simons term $\frac{i\kappa}{4\pi} \int \cA \wedge d\cA$ in the generating functional $F[\cA,\dots]$. It was argued in \cite{Closset:2012vp} that the fractional part of this term (in units of $\kappa$) is a physical observable. We will comment on this issue in section \ref{conc}.

Removing the divergence and the contact term, we are left with the parity-even result
\begin{align}
  \langle J_-^{(1)}(-q) J_+^{(1)} \rangle =
  -\frac{N\sin(\pi\lambda)}{16\pi\lambda}|q| \ed
\end{align}
This 2-point function is uniquely determined by conformal invariance and current conservation up to an overall coefficient, and we may therefore write
\begin{align}
  \langle J_\mu^{(1)}(-q) J_\nu^{(1)} \rangle =
  \frac{N \sin(\pi\lambda)}{\pi\lambda} \cdot
  \frac{q_\mu q_\nu - \delta_{\mu\nu} q^2}{16|q|} \ed
  \label{J1J1}
\end{align}
In appendix \ref{positivity} we verify that this result is consistent with reflection positivity.

\subsection{3-point Functions}
\label{3pt}

\subsubsection{$\left<J^{(0)}J^{(0)}J^{(0)}\right>$}

Evaluating the diagrams shown in Figure \ref{fig:JsJsJs} (a) using the exact $J^{(0)}$ vertex, we obtain
\begin{align}
  \langle J^{(0)}(-q) J^{(0)}(-q') J^{(0)}(-q'') \rangle &=
  -N \int \! \frac{d^3p}{(2\pi)^3} \,
  \text{Tr}\Big[ S(p+q)V(q,p)S(p)V(q'',p)S(p-q'')V(q',p)
  \notag \\ &\quad \qquad \qquad \qquad \quad \,
  +S(p+q)V(q,p)S(p)V(q',p)S(p-q')V(q'',p) \Big]
  \notag \\
  &= \frac{N\tan^2\!\left(\frac{\pi\lambda}{2}\right)}{2\pi\lambda} \ed
  \label{J0J0J0}
\end{align}

This is a pure contact term. For $\lambda=0$ this correlator vanishes since $J^{(0)}$ is parity-odd. For non-zero $\lambda$ the vanishing of this correlator was predicted in \cite{Giombi:2011kc} by computations in Vasiliev's gravity theory and by a perturbative calculation in field theory. This result is also consistent with the general analysis of Maldacena and Zhiboedov \cite{Maldacena:2012sf}.

\subsubsection{$\left<J^{(0)} J_+^{(1)} J_-^{(1)}\right>$}
Finally, using the $J^{(0)}$ and $J^{(1)}_{\pm}$ vertices in Figure \ref{fig:JsJsJs} (b), we obtain
\begin{align}
  \langle J^{(0)}(-q) J_+^{(1)}(-q') J_-^{(1)}(-q'') \rangle &= -N \int \frac{d^3p}{(2\pi)^3}
  \text{Tr}\Big[ S(p+q)V(q,p)S(p)V^-(q',p)S(p-q')V^+(q'',p) \notag \\
  &\quad \qquad\qquad\qquad\quad\,\,
  + S(p+q)V(q,p)S(p)V^+(q'',p)S(p-q'')V^-(q',p) \Big] \notag\\
  &= \frac{N}{4\pi\lambda}\frac{1}{1+e^{\pi i \hlambda}}
  \frac{q''^2 [ 1-e^{\pi i (\hlambda+\hlambda'')} ] +
  q'^2 [e^{\pi i \hlambda} - e^{-\pi i\hlambda'}] }{q' q''}
  \ed
\end{align}
This correlator has one parity-even and one parity-odd structure, which are conformally-invariant \cite{Giombi:2011rz}. The $\lambda$-even part (which corresponds to the parity-odd structure) is
\begin{align}
  \langle J^{(0)}(-q) J_+^{(1)}(-q') J_-^{(1)}(-q'') \rangle_{\lambda-\text{even}} &=
  \frac{iN\sin(\pi\lambda)}{8\pi\lambda}
  \left[\frac{|q'|}{q''}-\frac{|q''|}{q'}+
  |q|\left(\frac{1}{q'} - \frac{1}{q''}\right)\right] \ec
  \label{J0JpJmEven}
\end{align}
and the $\lambda$-odd part is
\begin{align}
  \langle J^{(0)}(-q) J_+^{(1)}(-q') J_-^{(1)}(-q'') \rangle_{\lambda-\text{odd}} &=
  \frac{N \sin^2\!\left(\frac{\pi\lambda}{2}\right)}{4\pi\lambda}
  \left[
  |q| \left(\frac{|q|}{q'q''}+\frac{|q'|}{qq''}+\frac{|q''|}{qq'}\right)
  -2
  \right] \ed
  \label{J0JpJmOdd}
\end{align}
The last term in the brackets is a contact term that we discard.

\section{Analysis of the Results}
\label{analysis}

Let us summarize our results, omitting contact terms. First, the 2-point functions.
\paragraph{\underline{Fermion 2-point functions:}}
\begin{align}
\langle J^{(0)} J^{(0)} \rangle &= \frac{4N\tan\!\left(\frac{\pi\lambda}{2}\right)}{\pi\lambda} \langle J^{(0)} J^{(0)} \rangle_{\text{fer.}} \ec \label{J0J0f}\\
\langle J^{(1)} J^{(1)} \rangle &= \frac{2N\sin\!\left(\pi\lambda\right)}{\pi\lambda} \langle J^{(1)} J^{(1)} \rangle_{\text{fer.}} \ed \label{J1J1f}
\end{align}
Here, $\langle\cdot\rangle_{\text{fer.}}$ are equal to one half times the same correlators in the theory of a single free Dirac fermion. They are obtained by setting $\lambda=0$ and $N=1/2$ in the results of section \ref{2pt}. Next, we write the normalized 3-point functions, defined by
\begin{align}
  \left<J^{(s_1)} J^{(s_2)} J^{(s_3)}\right>^{\text{norm.}} \equiv
  \frac{\left<J^{(s_1)} J^{(s_2)} J^{(s_3)}\right>}
  {\sqrt{|\left<J^{(s_1)} J^{(s_1)}\right>
  \left<J^{(s_2)} J^{(s_2)}\right>
  \left<J^{(s_3)} J^{(s_3)}\right>|}}
  \ed
\end{align}
\paragraph{\underline{Fermion 3-point functions:}}
\begin{align}
  \langle J^{(0)} J^{(0)} J^{(0)} \rangle^{\text{norm.}} &= 0
  \ec \label{J0J0J0f}\\
  \langle J^{(0)} J^{(1)} J^{(1)} \rangle^{\text{norm.}} &=
  \left[\frac{\pi\lambda}{4N\tan\!\left(\frac{\pi\lambda}{2}\right)}\right]
  ^{1/2}
  \langle J^{(0)} J^{(1)} J^{(1)} \rangle^{\text{norm.}}_{\text{fer.}}
  \notag\\
  &\quad + \text{sign}(\lambda) \cdot
  \left[\frac{\pi\lambda\tan\!\left(\frac{\pi\lambda}{2}\right)}{4N}\right]
  ^{1/2}
  \langle J^{(0)} J^{(1)} J^{(1)} \rangle^{\text{norm.}}_{\text{odd}}
  \ed \label{J0J1J1f}
\end{align}
Here, $\langle\cdot\rangle_{\text{fer.}}^{\text{norm.}}$ is again half the normalized correlator in the theory of a free Dirac fermion, and $\langle\cdot\rangle_{\text{odd}}^{\text{norm.}}$ is an independent structure that only appears in interacting theories. In fact $\langle J^{(0)} J^{(s)} J^{(s')} \rangle_{\text{odd}}$, for $s,s'>0$, is equal to the single conformally-invariant structure which appears in the same correlator in the critical bosonic vector model. We therefore define this structure to be half the normalized correlator in the theory of a single critical complex scalar. This last correlator can be obtained by taking $\lambda=0$ and $N=1/2$ in the result for critical bosonic Chern-Simons vector model, computed in \cite{Aharony:2012nh}.

As in the bosonic theory \cite{Aharony:2012nh}, we can use these results to determine the parameters $\tilde{N}$ and $\tilde{\lambda}$ of Maldacena and Zhiboedov \cite{Maldacena:2012sf}\footnote{In \cite{Maldacena:2012sf} only correlation functions of even-spin operators were computed, and we assume (as in \cite{Aharony:2012nh}) that these results extend naturally to a theory with operators of any integer spin.}. Since $J^{(1)}$ is canonically normalized\footnote{The current $J^{(1)}$ is canonically normalized in the sense that $[Q,\psi]=\psi$, where $Q$ is the $U(1)$ charge defined by the current.}, we can read off $\tilde{N}$ from the $J^{(1)}$ 2-point function. Then, $\tilde{\lambda}$ can be computed from $\langle J^{(0)} J^{(1)} J^{(1)} \rangle^{\text{norm.}}$. We find
\begin{align}
  \tilde{N} = 2N\frac{\sin(\pi\lambda)}{\pi\lambda} \ecq
  \tilde{\lambda} = \tan\!\left(\frac{\pi\lambda}{2}\right) \ed
  \label{micro}
\end{align}

\subsection{Comparison with Critical-Boson Theory}

We now determine the duality relation between the fermionic and critical bosonic theories. In the fermionic theory, let us denote $\lambda$ by $\lambda_f$ and $N$ by $N_f$. In \cite{Aharony:2012nh}, the variables $\tilde{\lambda}$ and $\tilde{N}$ were computed in terms of the parameters $\lambda_b$ and $N_b$ of the critical bosonic theory. They are given by $\tilde{N} = 2 N_b \sin(\pi \lambda_b) / \pi \lambda_b$ and $\tilde{\lambda} = - \cot(\pi\lambda_b/2)$. Note that this $\tilde{\lambda}$ has a minus sign compared with the one written down in \cite{Aharony:2012nh}. It comes from matching the conformal structures of the critical bosonic theory to the ones we are using in the fermionic theory\footnote{To see this, one can use the results of \cite{Maldacena:2012sf} (and the values of $\tilde{N}$ and $\tilde{\lambda}$ written above) to compute the even structure in $\langle J^{(0)} J^{(1)} J^{(1)} \rangle^{\mathrm{norm.}}$. This is done by first computing $\langle \hat{J}^{(0)} J^{(1)} J^{(1)} \rangle^{\mathrm{norm.}}$, where $\hat{J}^{(0)} = \tilde{\lambda} J^{(0)}$ is a parity-odd scalar operator. This is necessary because in \cite{Maldacena:2012sf} it is assumed that the dimension 2 scalar operator has odd parity. The result can then be verified to agree with the direct calculation of \cite{Aharony:2012nh}, eq. (70).}.

We equate the parameters of the critical bosonic theory with \eqref{micro},
\begin{align}
  \tilde{N} &= 2 N_f \frac{\sin(\pi \lambda_f)}{\pi \lambda_f}
  = 2 N_b \frac{\sin(\pi \lambda_b)}{\pi \lambda_b}
  \label{Nt} \ec \\
  \tlambda &= \tan\!\left( \frac{\pi\lambda_f}{2} \right)
  = -\cot\!\left( \frac{\pi\lambda_b}{2} \right)
  \label{lt} \ed
\end{align}
Notice that the 2-point functions of $J^{(1)}$ in the two theories change sign at $\lambda_f,\lambda_b=\pm 1$. Therefore, the theories are unitary only when $|\lambda_f|,|\lambda_b|<1$. Now, equation \eqref{lt} implies that $\lambda_f,\lambda_b$ have opposite signs, and that
\begin{align}
  \cos^2\! \left( \frac{\pi\lambda_f}{2} \right) =
  \sin^2\! \left( \frac{\pi\lambda_b}{2} \right)
  \quad\Longrightarrow\quad
  \cos\! \left( \frac{\pi|\lambda_f|}{2} \right) =
  \sin\! \left( \frac{\pi|\lambda_b|}{2} \right)
  \ed
\end{align}
The solution is $|\lambda_f| + |\lambda_b| = 1$, and we also see that $\sin(\pi\lambda_f) = -\sin(\pi\lambda_b)$. Using $\lambda=N/k$, equation \eqref{Nt} then gives
\begin{align}
  k_f = -k_b \ed
  \label{k-map}
\end{align}
Finally, we have
\begin{align}
  N_f = |k_b| - N_b \ed
  \label{N-map}
\end{align}

\section{The Critical Model}
\label{crit}

In this section we consider the critical fixed point of the fermion theory, the 3D Gross-Neveu model with Chern-Simons interactions. We will define this theory in the planar limit as the Legendre transform of the theory \eqref{action} with respect to the scalar operator $\bar{\psi}\psi$.\footnote{One way to think of the critical theory is as the UV fixed point at the end of an RG flow that starts with the regular theory, deformed by the operator $(\bar{\psi}\psi)^2$. However, since this deformation is irrelevant, it is difficult to make this description into a rigorous definition.} We implement this by introducing a coupling $\sigma \bar{\psi}\psi$, and adding a path integral over $\sigma$. The spectrum of the critical theory at large $N$ is identical to that of the regular theory, except that the scalar operator (which is now $\sigma$) has dimension $1+O(1/N)$.

For the sake of clarity, let us briefly review how to compute correlators
in the critical theory. Consider first the generating functional of connected correlators in the regular theory (let us call it $S_{\mathrm{eff.}}$), that is defined by
\begin{align}
  e^{-S_{\mathrm{eff.}}(\sigma,\cA)} = \int \! \cD \psi \, \cD \bar{\psi} \, \cD \! A \,
  e^{-S - \int \! d^3x \, \sigma \bar{\psi} \psi
  - \int \! d^3x \, J \cdot \cA} \ed
  \label{Seff}
\end{align}
Here, $\cA$ collectively denotes sources of currents $J^{(s)}$ with $s>0$. In previous sections we computed $S_{\mathrm{eff.}}$ at large $N$, up to cubic order in the sources.

To define the critical theory we make $\sigma$ dynamical by including a path integral over it, so that $S_{\mathrm{eff.}}$ becomes the effective action for $\sigma$. In the large $N$ limit --- where we include only planar contributions on the right-hand side of $\eqref{Seff}$ --- the effective action is proportional to $N$, and it is therefore the $\sigma$-1PI effective action of the theory\footnote{Beyond the large $N$ limit one has to include $1/N$ corrections to $S_{\text{eff.}}$ from non-planar diagrams in the regular theory. In addition, the 1PI action of $\sigma$ now receives contributions from $\sigma$ loops, so the critical and regular theories are no longer related by a Legendre transform.}. For example, the 2-point function of $\sigma$ is given by the inverse of the quadratic $\sigma$ term in $S_{\mathrm{eff.}}$, namely
\begin{align}
  G(q) \equiv \langle \sigma(-q) \sigma \rangle_{\mathrm{crit.}} =
  (- \langle J^{(0)} J^{(0)} \rangle_{\mathrm{non-crit.}})^{-1} =
  \frac{4\pi\lambda \cot\left(\frac{\pi\lambda}{2}\right)}{N} \frac{1}{|q|} \ed
  \label{sigma-prop}
\end{align}
The other 2-point functions, $\langle J^{(s)} J^{(s)} \rangle$ with $s>0$, are unchanged in the critical theory.

Before proceeding, we note that the critical theory admits a $\sigma^3$ deformation that is exactly marginal at large $N$. It is analogous to the ``triple-trace'' $(\bar{\phi}^i \phi_i)^3$ deformation of the bosonic theory. The effect of turning on this deformation is to change the effective action by $\delta S_{\mathrm{eff.}} = N \int \! d^3x \, \frac{\lambda_6}{3!} \sigma^3$. At large $N$, this interaction only affects the $\sigma$ 3-point function.

To compute general connected correlators that involve $\sigma$, one may take the Legendre transform of $S_{\mathrm{eff.}}$ with respect to $\sigma$; the result is the generating functional of the critical theory. Up to 3-point functions, we may equivalently compute a connected correlator by reading off the $\sigma$-1PI correlator from $S_{\mathrm{eff.}}$, and multiplying each $\sigma$ leg by its 2-point function \eqref{sigma-prop}.

We are now ready to compute the parameters $\tilde{N}_{\qb}$ and $\tilde{\lambda}_{\qb}$ of the critical model, which is a ``quasi-bosonic'' theory in the language of \cite{Maldacena:2012sf}. Since the 2-point function $\langle J^{(1)} J^{(1)} \rangle$ is unchanged, the result \eqref{micro} for $\tilde{N}$ still holds, and we have
\begin{align}
  \tilde{N}_{\qb} = 2N\frac{\sin(\pi\lambda)}{\pi\lambda} \ed
\end{align}
To extract $\tilde{\lambda}_\qb$ we follow the conventions of \cite{Maldacena:2012sf} and consider a scalar operator with even parity by defining $\hat{\sigma} = N \tilde{\lambda}_{\qb} \sigma$. The factor of $N$ is added so that $\hat{\sigma}$ correlators will be of order $N$ in the planar limit.
Using \eqref{sigma-prop} and \eqref{J0JpJmOdd}, we find that
\begin{align}
  \langle \hat{\sigma} J^{(1)} J^{(1)} \rangle_{\mathrm{crit.~fer.,~even}}^{\mathrm{norm.}}
  = - \frac{\mathrm{sign}(\tilde{\lambda}_\qb \lambda)}{2}
  \left[ \frac{\pi\lambda \tan \left( \frac{\pi\lambda}{2} \right)}{N} \right]^{1/2}
  \langle J^{(0)} J^{(1)} J^{(1)} \rangle_{\mathrm{free~bos.}}^{\mathrm{norm.}}
  \ed
\end{align}
On the left-hand side we have the parity-even structure of the critical fermion model.
The correlator on the right-hand side is one-half the normalized 3-point function in the theory of a free complex scalar. Using the results of \cite{Maldacena:2012sf}, one can see that the right-hand side of this equation must have a positive coefficient, and therefore $\mathrm{sign}(\tilde{\lambda}_\qb) = - \mathrm{sign}(\lambda)$. Completing the calculation as in section \ref{analysis}, we find that
\begin{align}
  \tilde{\lambda}_\qb = - \cot \left( \frac{\pi\lambda}{2} \right) \ed
  \label{tlambda}
\end{align}
These results mirror those of the critical bosonic model \cite{Aharony:2012nh}. In terms of $k$ and $N$, the duality map is exactly the same as the one relating the regular fermion and the critical boson, appearing in \eqref{k-map} and \eqref{N-map}.

Finally, let us determine the duality map for the triple-trace coupling\footnote{We are grateful to S. Giombi for reminding us of this point.}. To this end we compare the 3-point function of $\hat{\sigma}$ with that of $J^{(0)}$ in the regular bosonic theory. For the 3-point function of $\hat{\sigma}$, using \eqref{J0J0J0}, \eqref{sigma-prop} and \eqref{tlambda} we obtain
\begin{align}
  \langle \hat{\sigma}(-q) \hat{\sigma}(-q') \hat{\sigma}(-q'') \rangle &= N^3 \tilde{\lambda}_\qb^3 \langle \sigma(-q) \sigma(-q') \sigma(-q'') \rangle \notag\\
  &= N^3 \tilde{\lambda}_\qb^3 G(q) G(q') G(q'') \left[ -N \lambda_6^f - \langle J^{(0)}(-q) J^{(0)}(-q') J^{(0)}(-q'')\rangle \right] \notag\\
  &= 32 N \pi^2 \lambda^2 \cot^4\left(\frac{\pi\lambda}{2}\right) \left[ 1 + 2\pi\lambda\lambda_6 \cot^2\left(\frac{\pi\lambda}{2}\right)\right] \frac{1}{|q||q'||q''|} \ed
\end{align}
Normalizing this correlator by the 2-point function \eqref{sigma-prop}, equating to the corresponding bosonic correlator $\langle J^{(0)} J^{(0)} J^{(0)} \rangle^{\mathrm{norm.}}$ that was computed in \cite{Aharony:2012nh}, and applying the duality map for $k$ and $N$, we obtain the relation between the triple-trace couplings in the two theories,
\begin{align}
\lambda_6^b = 8\pi^2(1-|\lambda_f|)^2\left( 3 - 8\pi \lambda_f \lambda_6^f \right) \ed
\end{align}

\section{Conclusions}
\label{conc}

In this work we obtained all the planar 3-point functions in the fermionic Chern-Simons vector models, by computing several correlators exactly and using the results of \cite{Maldacena:2012sf}. Using this result we completed the duality map of these theories to the bosonic Chern-Simons vector models, and verified that it works for all values of the 't Hooft coupling and of the triple-trace coupling.

Since we worked in momentum space, the results of our computation include the contact terms that appear in the 2- and 3-point functions. Contact terms are usually not universal, but in \cite{Closset:2012vp} it was noted that certain contact terms, specifically those in $\langle J^{(1)} J^{(1)} \rangle$ and $\langle J^{(2)} J^{(2)} \rangle$, are in fact physical observables (up to integer shifts).
One can therefore use these terms to further test the duality. It would also be interesting to understand if contact terms in other 2-point functions are physical, and whether they are constrained by high-spin symmetries. Moreover, the coefficients of these terms map under the $AdS$/CFT correspondence to the bulk $\theta$-angle and its higher-spin analogs in Vasiliev's theory. Since these terms do not affect the bulk equations of motion, they were not known previously, and it will be interesting to analyze them.

There are several issues one has to take into account when trying to match these contact terms under the duality between vector models\footnote{We thank Zohar Komargodski and Nathan Seiberg for discussions on this issue.}. First, the gauge and topological $U(1)$ currents can mix and one has to determine the combinations which map to each other under the duality. While this effect cannot influence correlators at separated points, it will in general affect the contact terms. Note that the contact term in $\langle J^{(2)} J^{(2)} \rangle$ does not suffer from such an ambiguity. Moreover, since our gauge and regulator break Lorentz invariance the correlators may contain Lorentz-violating contact terms in addition to the ones written down in \cite{Closset:2012vp}.

It is currently not known how the results of \cite{Maldacena:2012sf} extend to supersymmetric theories. It will be interesting to understand this, both using the current algebra techniques of \cite{Maldacena:2012sf}, and by directly computing the correlators as we did here. For instance, it will be interesting to see how this duality plays out in theories with $\mathcal{N}=1$ supersymmetry, as there are no similar known dualities in this case. We note that the form of the ``bosonization'' duality obtained in \cite{Aharony:2012nh} is quite similar to Giveon-Kutasov duality \cite{Giveon:2008zn} of $\cN=2$ supersymmetric theories, or one of its ``chiral'' versions \cite{Benini:2011mf}. It would be interesting to understand if the non-supersymmetric duality of this paper follows from the supersymmetric one. Analyzing $\cN=6$ theories in these methods would also be interesting, in particular given the relation between Vasiliev's theory and string theory proposed in \cite{Chang:2012kt}.

So far the tests of this duality mostly follow from the large $N$ high-spin symmetries of Chern-Simons vector models\footnote{The fact that the duality transformation at large $N$ is the same as level-rank duality, and the matching of mass deformations done in \cite{Aharony:2012nh}, do not follow from the symmetries as far as we know.}. It will therefore be interesting to find more evidence for the bosonization duality which does not follow only from symmetry considerations. One direction is to attempt to perform large $N$ computations of partition functions on various manifolds, or to consider correlators of Wilson loops, as these are not entirely determined from symmetries as far as we know.

One such possibility is to compute the thermal free energy at large $N$. This was done in \cite{Giombi:2011kc,Jain:2012qi}, but the results were found to be inconsistent with existing (conjectured) dualities.
One possible reason for the discrepancy, already mentioned in \cite{Jain:2012qi,Aharony:2012nh}, is that the holonomy of the gauge field along the thermal cycle might be non-trivial (as assumed in these computations). In \cite{AGGMY} it is shown that, indeed, when one carefully accounts for the holonomy the dualities hold.

It will be very interesting to find evidence for the duality beyond the large $N$ limit. At this point, it is not yet clear what are the high-spin symmetry constraints on non-planar correlators, or on planar $n$-point functions with $n>3$. One may try to understand whether such constraints exist and whether they are as strong as the ones on planar 3-point functions. We hope to return to these interesting issues in the future.

\subsection*{Acknowledgments}
\label{s:acks}

We would like to thank Ofer Aharony, Chi-Ming Chang, Simone Giombi, Zohar Komargodski, Juan Maldacena, Nathan Seiberg and Xi Yin for many useful discussions on related issues.
We would especially like to thank Ofer Aharony for his support in this work, and for his helpful comments on early versions of this paper.

\appendix

\section{Reflection Positivity}
\label{positivity}

In this appendix we briefly review the argument for the positivity of 2-point functions in a theory with a non-real action (as is the case with \eqref{action}), and we show what positivity implies for the 2-point functions we compute in section \ref{2pt}.

Consider a field theory in $d$-dimensional Euclidean space, with fields
collectively denoted by $\Phi$ and an action $S$. Suppose we have a
``parity'' transformation $\cP$ that satisfies $\cP^2=1$ and splits the space into two components, with a boundary that is given by $x=\cP x$. Let us further assume that $\cP$ leaves the path integral measure $\cD\Phi$ invariant, and that it acts as complex conjugation on the action, in the sense that
\begin{align}
  S(\cP \Phi) = S^*(\Phi) \ed
\end{align}
Note that $\cP$ is not a symmetry if the action is not real. Let us denote the fields, restricted to either space component, by $\Phi_\pm$, and the fields on the boundary by $\Phi_0$. Now, suppose we have a local operator $\cO$ that obeys $\cP \cO(x) = \cO(\cP x)$. We can then show that the 2-point function $\langle \cO \cO^\dagger \rangle$ is positive when evaluated at separated points related by parity. Indeed,
\begin{align}
  \langle \cO(x) \cO^\dagger(\cP x) \rangle &=
  \int \! \cD\Phi \, e^{-S(\Phi)} \cO(x) O^\dagger(\cP x)
  \notag \\
  &= \int \! \cD\Phi_0 \,
  \int \! \left[ \cD\Phi_+ \right]_{\left.\Phi\right|_{\dho}=\Phi_0}
  e^{-S(\Phi_+)} \cO(x)
  \int \! \left[ \cD\Phi_- \right]_{\left.\Phi\right|_{\dho}=\Phi_0}
  e^{-S(\Phi_-)} O^\dagger(\cP x)
  \notag \\
  &= \int \! \cD\Phi_0 \,
  \left|
  \int \! \left[ \cD\Phi_+ \right]_{\left.\Phi\right|_{\dho}=\Phi_0}
  e^{-S(\Phi_+)} \cO(x)
  \right|^2 \ge 0 \ec
  \label{pos}
\end{align}
where in the last step we made the change of variables $\Phi_- \to \cP \Phi_- = \Phi_+$. Similarly, for an operator satisfying $\cP \cO(x) = -\cO(\cP x)$, the correlator $\langle \cO \cO^\dagger \rangle$ is negative.

For the action \eqref{action}, which is purely imaginary, we can define a parity transformation which has the desired properties:
\begin{align}
  x^3 \to -x^3 \ecq
  \psi \to \gamma^3 \psi \ecq
  \bar{\psi} \to \bar{\psi} \gamma^3 \ecq
  A_\mu \to (-1)^{\delta_{\mu 3}} A_\mu \ed
\end{align}
Under this $\cP$ we have $J^{(0)} \to J^{(0)}$ and $J^{(1)}_\pm \to - J^{(1)}_\pm$.
Reflection positivity then implies that
\begin{align}
  \langle J^{(0)}(x) J^{(0)}(\cP x) \rangle &\ge 0 \ecq
  \\
  \langle J^{(1)}_+(x) J^{(1)}_-(\cP x) \rangle
  &= - \langle J^{(1)}_+(x) J^{(1)\dagger}_+(\cP x) \rangle
  \ge 0 \ed
\end{align}
Using the known conformal structures \cite{Osborn:1993cr} for these correlators, we find that the general 2-point functions are given by
\begin{align}
  \langle J^{(0)}(x) J^{(0)}(0) \rangle &= \frac{\tau_0}{|x|^4} \ec \\
  \langle J^{(1)}_\mu(x) J^{(1)}_\nu(0) \rangle &= \tau_1
  \frac{\delta_{\mu\nu} x^2 - 2 x_\mu x_\nu}{|x|^6} \ec
\end{align}
with positive $\tau_0$ and $\tau_1$. It is easy to verify that this conclusion is consistent with our momentum-space results \eqref{J0J0} and \eqref{J1J1}.


\end{document}